\documentclass[preprint,12pt]{elsarticle}
\usepackage{graphicx}
\usepackage{colordvi}
\journal{J. Phys. B}
\begin{document}

\begin{frontmatter}
\title{Weakly bound solitons and two-soliton molecules in dipolar Bose-Einstein condensates}
\author[a]{B. B. Baizakov\corref{*}}
\ead{baizakov@uzsci.net} \cortext[*]{Corresponding author.}
\author[b]{S. M. Al-Marzoug}
\author[c]{U. Al Khawaja}
\author[b]{H. Bahlouli}

\address[a]{Physical-Technical Institute, Uzbek Academy of
Sciences, 100084, Tashkent, Uzbekistan}
\address[b]{Physics Department, King Fahd University of Petroleum and Minerals, \\
and Saudi Center for Theoretical Physics, Dhahran 31261, Saudi
Arabia}
\address[c]{Physics Department, United Arab Emirates University, P.O. Box
15551, Al-Ain, United Arab Emirates}

\begin{abstract}
Analytic expressions have been derived for the interaction potential
between dipolar bright solitons and the binding energy of a
two-soliton molecule. The properties of these localized structures
are explored with a focus on their behavior in the weakly bound
regime, with a small binding energy. Using the variational approach
a coupled system of ordinary differential equations for the
parameters of a soliton molecule is obtained for the description of
their evolution. Predictions of the model are compared with
numerical simulations of the governing nonlocal Gross-Pitaevskii
equation and good qualitative agreement between them is
demonstrated.
\end{abstract}


\end{frontmatter}

\section{Introduction}

The realization of Bose-Einstein condensation (BEC) of atomic
species with large natural magnetic moments, such as chromium
\cite{griesmaier2005,beaufils2008}, dysprosium
\cite{lu2011,tang2015}, erbium \cite{aikawa2012} and heteronuclear
dipolar quantum mixtures of erbium and dysprosium atoms
\cite{trautmann2018} has initiated new directions of research in the
field of ultra-cold quantum gases. The physical properties of
dipolar condensates are essentially different~\cite{lahaye2009} as
compared to other members of the BEC family with contact atomic
interactions \cite{pethick-book,pitaevskii-book}. The differences
originate from the long-range and anisotropic dipole-dipole
interaction potential between atoms in the gas, which decay with
distance as $\sim 1/x^3$. In contrast to this, the atoms in
non-dipolar BEC interact only when they are at the same spatial
location, the so called contact interactions. The effects of the
anisotropic and long range dipolar atomic interactions were observed
in diverse forms, such as in intriguing collapse mechanism
\cite{lahaye2008}, emergence of a roton-like branch in the
excitation spectrum \cite{santos2003,petter2018}, modification of
superfluid properties~\cite{wilson2010,ticknor2011}, peculiar
density fluctuations in the ground state \cite{zaman2010}, Faraday
patterns sensitive to the roton mode \cite{nath2010} and other novel
many-body effects in dipolar quantum gases
\cite{baranov2008,yukalov2016}. The properties of dipolar BEC in
quasi-2D traps were analyzed in \cite{baillie2015} and a number of
analytic approximations for the condensate excitation spectrum were
introduced.

Among a rich variety of nonlinear wave phenomena in dipolar BEC a
particularly interesting one is the existence of stable isotropic
and anisotropic 2D solitons
\cite{pedri2005,tikhonenkov2008,tikhonenkov2008a}, given the fact
that multi-dimensional solitons in conventional BEC with contact
atomic interactions are unstable against collapse or decay. In
quasi-1D BEC with competing dipolar and local interactions bright
soliton families were found in free space and in the presence of
optical lattices~\cite{cuevas2009}. In this work and subsequent
papers \cite{young2011,eichler2012,umarov2016} by analyzing
different regimes of soliton collisions the regions in the parameter
space where colliding solitons either merge or pass through each
other were investigated. Dynamics and head-on collisions of solitons
in dipolar BEC confined to 2D and 3D traps were investigated by
variational approach and numerical simulations \cite{adhikari2014}.
The transmission, reflection and splitting regimes for a dipolar
soliton moving in a parabolic trap with a barrier potential at its
center were also studied \cite{abdullaev2012}. The existence,
stability and collisions of dipolar bright solitons, showing a
number of novel features, were analyzed \cite{edmonds2017}. The
dipolar BEC with repulsive nonlinearity supports dark solitons,
appearing as a traveling local reduction of the field amplitude.
Distinctive features of dark solitons in dipolar BEC, such as
interaction sensitive oscillations in external potentials, were
reported \cite{bland2017}.

During the last few years the existence, stability and dynamics of
soliton bound states in dipolar BEC, so called \textit{soliton
molecules}, have attracted considerable interest. The possibility of
molecular type of interaction between two solitons, reflected
through the fact that they attract each-other at long distances and
repel at short distances was theoretically studied
\cite{nath2007,lakomy2012}. In these models individual solitons,
forming the bound state, reside in separate layers. Numerical
analysis of soliton bound states in quasi-2D dipolar BEC was
presented \cite{lashkin2007}. The vibration spectrum of a
two-soliton molecule in dipolar BEC confined to a quasi-1D trap was
studied in \cite{turmanov2015}, while the conditions when two
solitons can form a stable bound state has been analyzed using the
variational approach and numerical simulations \cite{baizakov2015}.
Recently similar bound states of dark solitons in dipolar BEC were
reported in \cite{pawlowski2015}, where the shape of the interaction
potential between two dark solitons was numerically generated. The
emission of phonons by colliding dark solitons and their
simultaneous acceleration was another important finding in that
work. The existence, stability and collision dynamics of dark
solitons in quasi-1D dipolar BEC along with formation of bound
states of dark solitons were investigated \cite{bland2015}.

Nowadays research on dipolar quantum gases in general, and dipolar
solitons in particular, is actively pursued due to novel physical
phenomena observed in dipolar BEC
\cite{ferrier-barbut2016,kadau2016,schmitt2016} and many promising
practical applications \cite{carr2009,helm2015}. For instance, the
authors of a recent work \cite{sapina2013} have demonstrated that
oscillations of a dipolar condensate near the superconducting
surface (atom chip) can induce eddy currents in it, which act back
on the BEC. The interaction of a dipolar BEC with a superconductor
via eddy currents provides a new physical mechanism for the coupling
between these two macroscopic quantum systems. The eddy currents can
change the oscillation frequency of the condensate moving near the
superconducting surface and excite collective modes in
BEC~\cite{sapina2014}.

Despite the good progress achieved in understanding the physics of
dipolar condensates and propagation of nonlinear waves in these
media, there are many open issues awaiting their clarification. In
particular, when the short-range contact interactions and long-range
dipolar interactions between atoms nearly balance each-other an
additional term in the equation, the so called beyond mean field
term, becomes important \cite{lima2011,edler2017}. This term
originating from quantum fluctuations is responsible for the
existence of quantum droplets, a novel ultra dilute liquid state of
matter \cite{ferrier-barbut2016,kadau2016,schmitt2016}. The
existence regimes, stability and dynamics of quantum droplets in
dipolar BEC are among the actively explored topics in contemporary
physics of ultracold gases.

Our objective in this work is to derive the analytic form of the
interaction potential between two bright solitons in dipolar BEC and
estimate the binding energy of a two-soliton molecule. Special
attention will be given to investigation of weakly bound states of
these localized structures with a small binding energy. Next we
explore the dynamics of bright solitons and two-soliton molecules
under slowly varying strength of dipolar interactions. We address
the problem analytically using a variational approach (VA), and
numerically by solving the governing nonlocal Gross-Pitaevskii
equation (GPE). The resulting equations will allow to reveal the
features of dipolar solitons and two-soliton molecules in the weakly
bound regime, when the competing short-range contact interactions
($q<0$) and long-range dipolar interactions ($g>0$) nearly balance
each-other. However, despite the smallness of the net attractive
interactions, the above mentioned beyond mean field effects are
supposed to be still negligible, and omitted from the governing
equation.

\section{Model equations and variational approach}

We consider the dynamics of single solitons and two-soliton
molecules in a dipolar BEC confined to a quasi-1D trap. The
governing equation of our model is the one-dimensional nonlocal GPE
for the mean-field wave function of the condensate $\psi(x,t)$
\begin{equation} \label{gpe}
i\psi_t+\frac{1}{2}\psi_{xx} + q|\psi|^2\psi+g\psi(x,t)\int_{-\infty
}^{+\infty} R(|x-x'|)\ |\psi (x',t)|^2 \ dx' =0,
\end{equation}
where the subscripts denote partial derivatives. The dimensionless
quantities, entering this equation are scaled using the frequency of
the radial confinement $\omega_{\bot}$, atomic mass $m$ and radial
harmonic oscillator length $\l_{\bot} = \sqrt{\hbar/m\omega_{\bot}}$
as follows: time $t \rightarrow t \omega_{\bot}$, space $x
\rightarrow x/l_{\bot}$, wave function $\psi \rightarrow
\sqrt{2a_{s0}}\psi$, the coefficient of contact interactions $q
\rightarrow a_s/|a_{s0}|$, the coefficient of long-range dipolar
interactions $g \rightarrow a_d/|a_{s0}|$. The background values of
the atomic scattering length $a_{s0}$ and the length scale of
dipolar interactions $a_d=md\,^2/\hbar^2$ are supposed to be known
characteristics of the condensate. Although, in experiments both the
atomic scattering length $a_s$ and dipole moment of atoms $d$ can be
varied through external magnetic, electric or optical fields.

The response function $R(x)$ in the integral term of Eq. (\ref{gpe})
characterizes the degree of non-locality of the medium. It shows how
strongly the properties of the medium at a given location depend on
the properties of its neighborhood. An analytic expression for the
response function of a quasi-1D dipolar BEC was obtained in a single
mode approximation \cite{sinha2007}.
\begin{equation}\label{kernel-ss}
R(x) = (1+2x^{2}) \, \exp(x^2) \, \mathrm{erfc}(|x|)-2\pi^{-1/2}|x|.
\end{equation}
While relevant analytic calculations are complicated it is worth
mentioning that the response function is characterized by a cusp at
$x=0$. For practical reasons, however, another function was proposed
in~\cite{cuevas2009}, which behaves smoothly at the origin and is
more convenient for analytical treatment
\begin{equation}\label{kernel}
R(x) = (\sigma^2 x^2+1)^{-3/2}.
\end{equation}
The value of the cutoff parameter $\sigma=\pi^{1/2}$ was found from
the condition of equal areas beneath the curves of the single mode
kernel and the one given by Eq. (\ref{kernel}). A very good
correspondence between the two response functions was noted (see
Fig. 1 in Ref. \cite{cuevas2009}). Hence, our variational
calculations will be based on the response function given by Eq.
(\ref{kernel}).

For an arbitrary external potential and nonlocal term Eq.
(\ref{gpe}) does not possess analytic solution. The variational
approach \cite{anderson1983,malomed2002} is frequently used to study
the dynamics of solitons governed by this type of equations. To
develop the VA for our model we note that Eq. (\ref{gpe}) can be
generated from the following Lagrangian density
\begin{eqnarray} \label{lagden}
{\cal L} &=& \frac{i}{2}(\psi \psi^{\ast}_t - \psi^{\ast}\psi_t) +
\frac{1}{2} |\psi_x|^2 - \frac{q}{2}|\psi|^4 -
\nonumber \\
 & & \frac{g}{2}|\psi(x,t)|^2 \int \limits_{-\infty}^{\infty} R(|x-x'|) |\psi(x',t)|^2 dx'.
\end{eqnarray}
The success of the VA strongly depends on the right choice of the
trial function. On one hand the selected function should closely
match the shape of the localized state (soliton or molecule), and on
the other hand it should lead to analytically tractable integrals
associated with this Lagrangian. Below we develop the suitable
models for a single soliton and two-soliton molecule using the
Gauss-Hermite trial functions.

\subsection{Variational approach for single solitons}

To derive variational equations for a single soliton moving in the
external potential we use the Gaussian trial function
\begin{equation}\label{ansatz1}
\psi(x,t)=A \exp \left[ -\frac{(x-\xi)^2}{2a^2} + i b (x-\xi)^2 + i
v (x-\xi) + i\phi \right],
\end{equation}
where $A(t),a(t),b(t),\xi(t),v(t),\phi(t)$ are time-dependent
variational parameters, meaning the amplitude, width, chirp
parameter, position of the center-of-mass, velocity and phase of the
soliton, respectively. The velocity is defined as a time derivative
of the soliton's center-of-mass position $v=\xi_t$. The norm $N =
A^2 a \sqrt{\pi}$, which is a conserved quantity of the governing
equation, is proportional to the number of atoms in the condensate.

Substitution of Eq. (\ref{kernel}) and Eq. (\ref{ansatz1}) in the
Lagrangian density (\ref{lagden}), followed by spacial integration,
$L = \int_{-\infty}^{\infty} {\cal L} dx $, leads to the effective
Lagrangian
\begin{equation}
\frac{L}{N}=\frac{1}{2}a^2b_t-\frac{1}{2}\xi_t^2+\phi_t+
\frac{1}{4a^2}+a^2b^2 - \frac{q N}{2\sqrt{2\pi}a} -\frac{g N}{2
\sqrt{2}\sigma a} {\cal U}\left(\frac{1}{2},0,\frac{1}{2\sigma^2
a^2}\right),
\end{equation}
where ${\cal U}({\rm a},{\rm b},{\rm c}) = \frac{1}{\Gamma({\rm
a})}\int_0^{\infty} e^{-{\rm c} t} t^{{\rm a}-1} (t+1)^{{\rm b}-{\rm
a}-1}dt$ is the confluent hypergeometric function \cite{abramowitz},
with $\Gamma({\rm a})$ being the gamma function. Now using the
associated Euler-Lagrange equations $d/dt(\partial L/\partial \nu_t)
-\partial L/\partial \nu = 0$ for the variational parameters
$\nu~\rightarrow~a,~b,~\xi,~\phi$ one can derive the following
equation for the width
\begin{equation}\label{att1}
a_{tt}=\frac{1}{a^3}-\frac{q N}{\sqrt{2\pi}a^2}+\frac{g
N}{\sqrt{2}\sigma}\frac{\partial F_s(a,\sigma)}{\partial a},
\end{equation}
where the non-locality function $F_s(a,\sigma)$ is given by
\begin{equation}\label{Fs}
F_s(a,\sigma)=\frac{1}{a} \, {\cal
U}\left(\frac{1}{2},0,\frac{1}{2\sigma^2 a^2}\right).
\end{equation}
The stationary width $a_0$ and amplitude $A_0=\left[N/(a_0
\sqrt{\pi})\right]^{1/2}$ of the soliton with a given norm $N$ can
be found from the fixed point of Eq.~(\ref{att1}).

\begin{figure}[htb]
\centerline{
\includegraphics[width=6cm,height=6cm,clip]{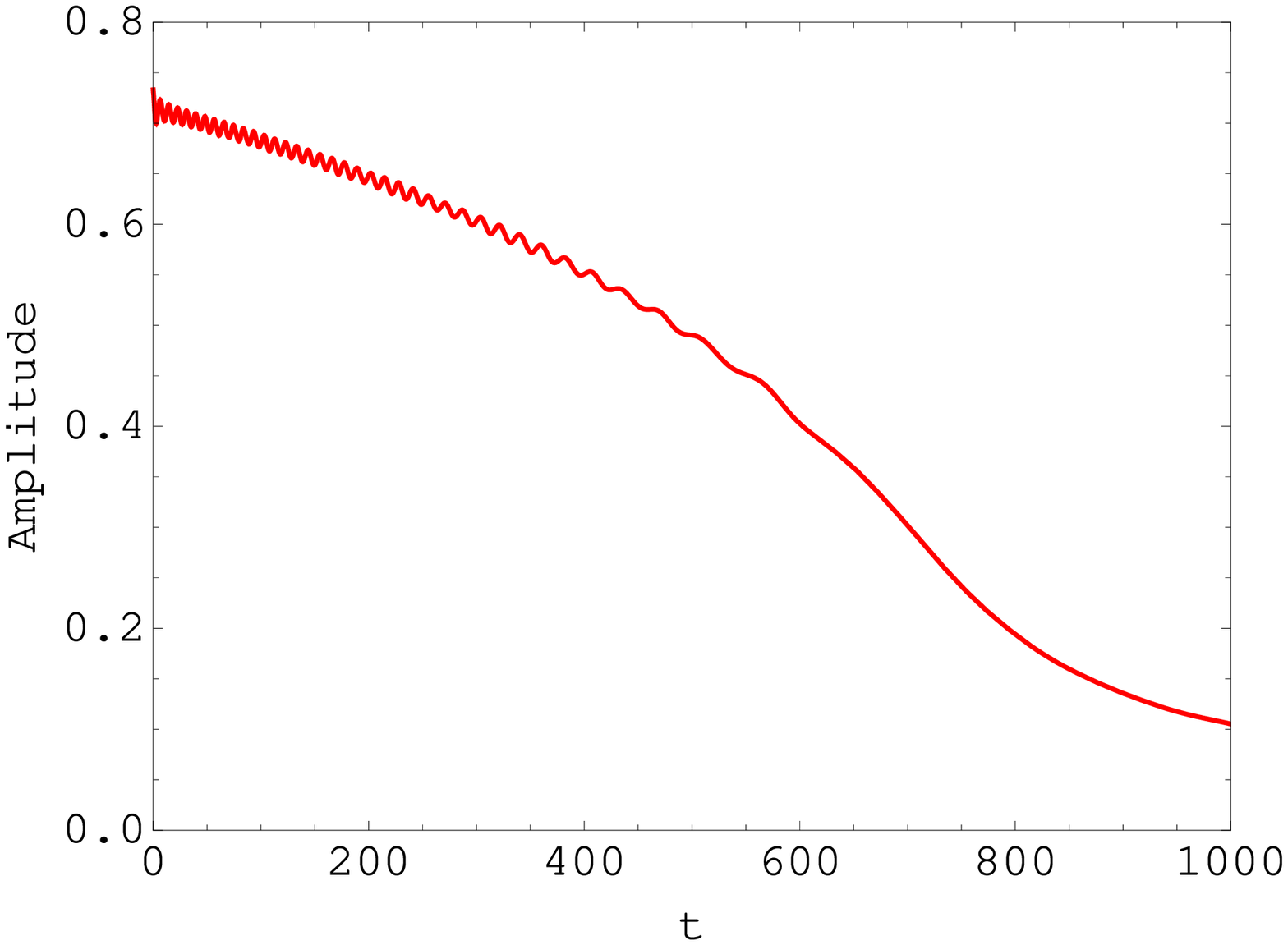}\qquad
\includegraphics[width=6cm,height=6cm,clip]{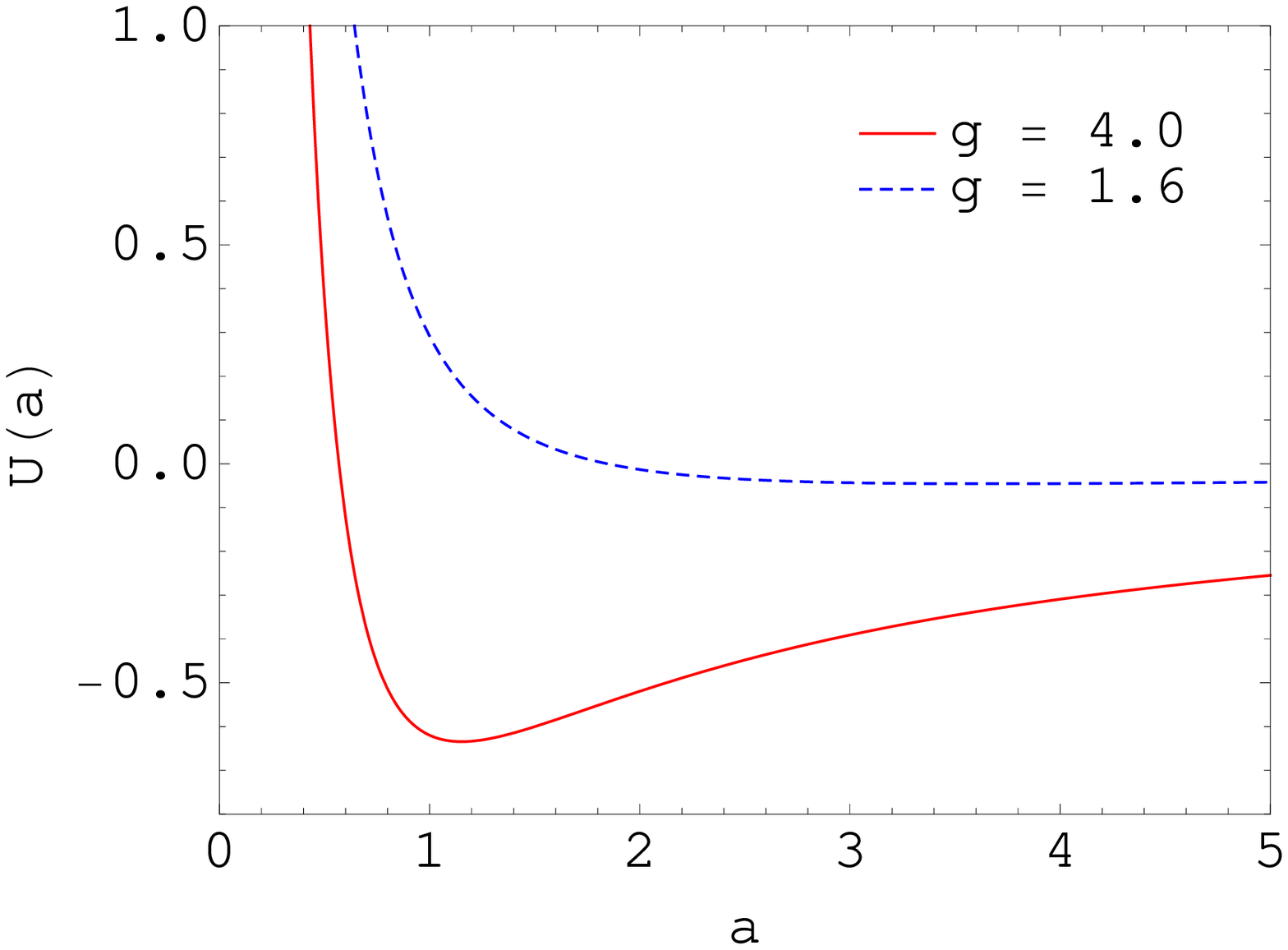}}
\caption{Left panel: The soliton amplitude according to numerical
solution of the GPE~(\ref{gpe}) with the VA predicted initial
condition and linearly decreasing strength of dipolar interactions
$g(t)=g_0 (1-t/t_{end})$ for parameters $N=1$, $A=0.7$, $a=1.15$,
$q=-1$, $g_0 = 4, \, t_{end}=1000$. Until $t\simeq 600$ which
corresponds to $g_{cr}=1.6$ the amplitude decreases in an
oscillatory manner, after that oscillations cease, indicating to
disintegration of the soliton. Right panel: The potential Eq.
(\ref{pot1}) for the initial ($g=4$) and critical ($g=1.6$) values
of the strength of dipolar interactions. The local minimum
disappears below the critical value, therefore self-bound localized
states do not exist in this case. } \label{fig1}
\end{figure}

The Eq. (\ref{att1}) has a formal analogy with the equation of
motion for a unit mass particle moving in the anharmonic potential
\begin{equation}\label{pot1}
a_{tt} = - \frac{\partial U}{\partial a}, \qquad U(a) = \frac{1}{2
a^2}-\frac{q N}{\sqrt{2 \pi} a} - \frac{g N}{\sqrt{2} \sigma}
F_s(a,\sigma).
\end{equation}
The presence of a local minimum in this potential for a given set of
parameters indicates the existence of a self-bound localized state,
which is also called soliton, despite the non-integrability of the
governing Eq. (\ref{gpe}). In Fig. \ref{fig1} we illustrate the
numerical experiment with Eq. (\ref{gpe}) when the VA predicted wave
profile Eq. (\ref{ansatz1}) is introduced as initial condition, and
the strength of the dipolar interactions is linearly decreased with
time. Since the variational solution is approximate one, at the
beginning the amplitude while decreasing oscillates at a high
frequency, which is an evidence for a strongly bound soliton. At
some critical value of the dipolar strength $g=g_{cr}$ the soliton
disintegrates, and the local minimum in the potential Eq.
(\ref{pot1}) disappears. Below this critical value of the strength
of dipolar interaction the net nonlinearity in Eq. (\ref{gpe})
becomes repulsive, and the system does not support self-bound bright
localized states. Throughout the paper we consider the repulsive
contact interactions with $q=-1$.

\subsection{Variational approach for two-soliton molecules}

As a trial function for two-soliton molecules we select the
Gauss-Hermite function proposed in \cite{pare1999,feng2004}
\begin{equation}\label{ansatz2}
\psi(x,t)=A (x-\xi) \exp \left[ -\frac{(x-\xi)^2}{2a^2} + i b
(x-\xi)^2 + i v (x-\xi) + i\phi \right],
\end{equation}
where the meaning of the variational parameters are similar to those
of a single soliton considered above. It should be noted that Eq.
(\ref{ansatz2}) adequately describes the properties of two-soliton
molecules when the solitons move near their equilibrium positions.
At large separations of solitons this waveform deviates from the
superposition of two anti-phase solitons, leading to a lower
accuracy of the method.

\begin{figure}[htb]
\centerline{
\includegraphics[width=6cm,height=6cm,clip]{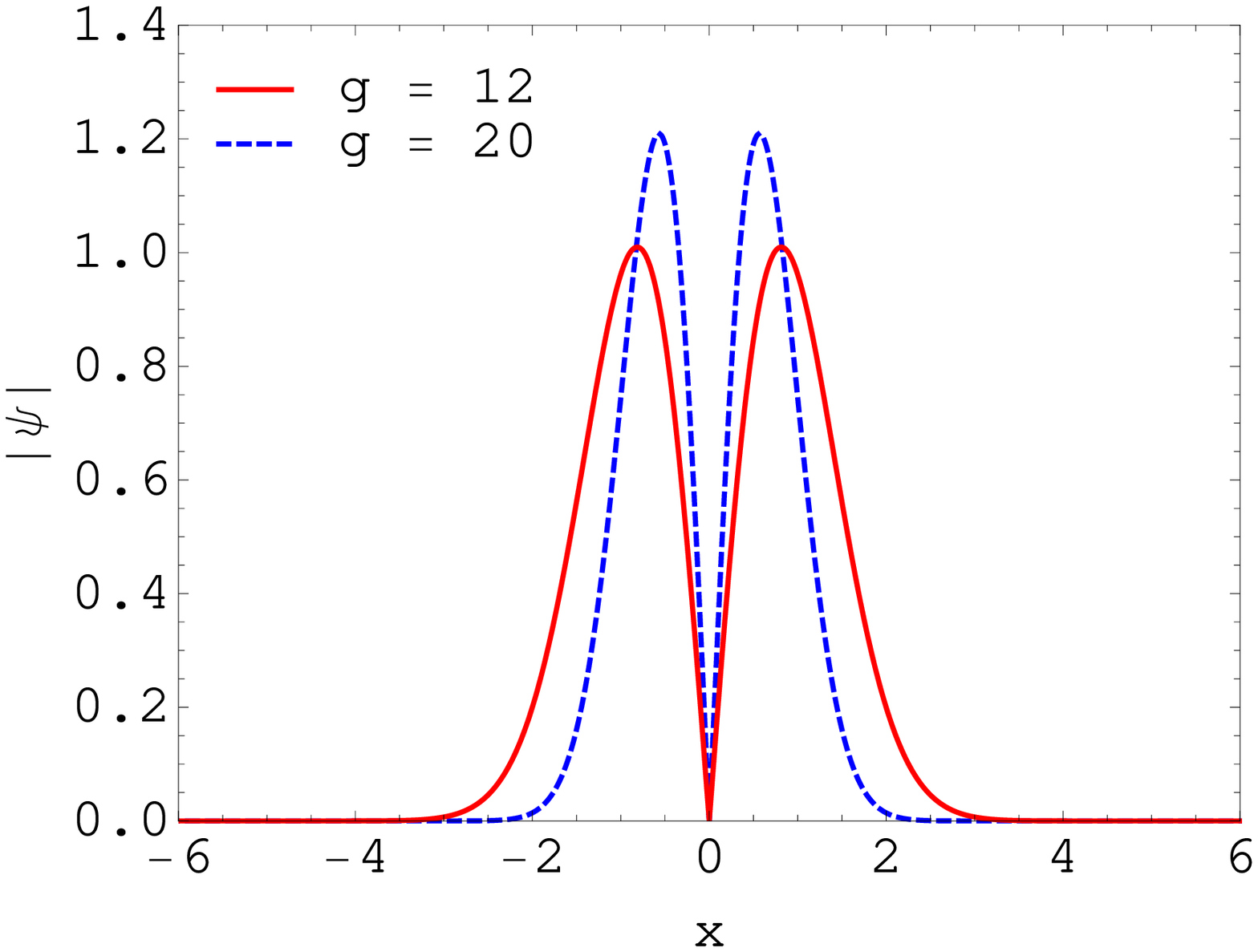}\qquad
\includegraphics[width=6cm,height=6cm,clip]{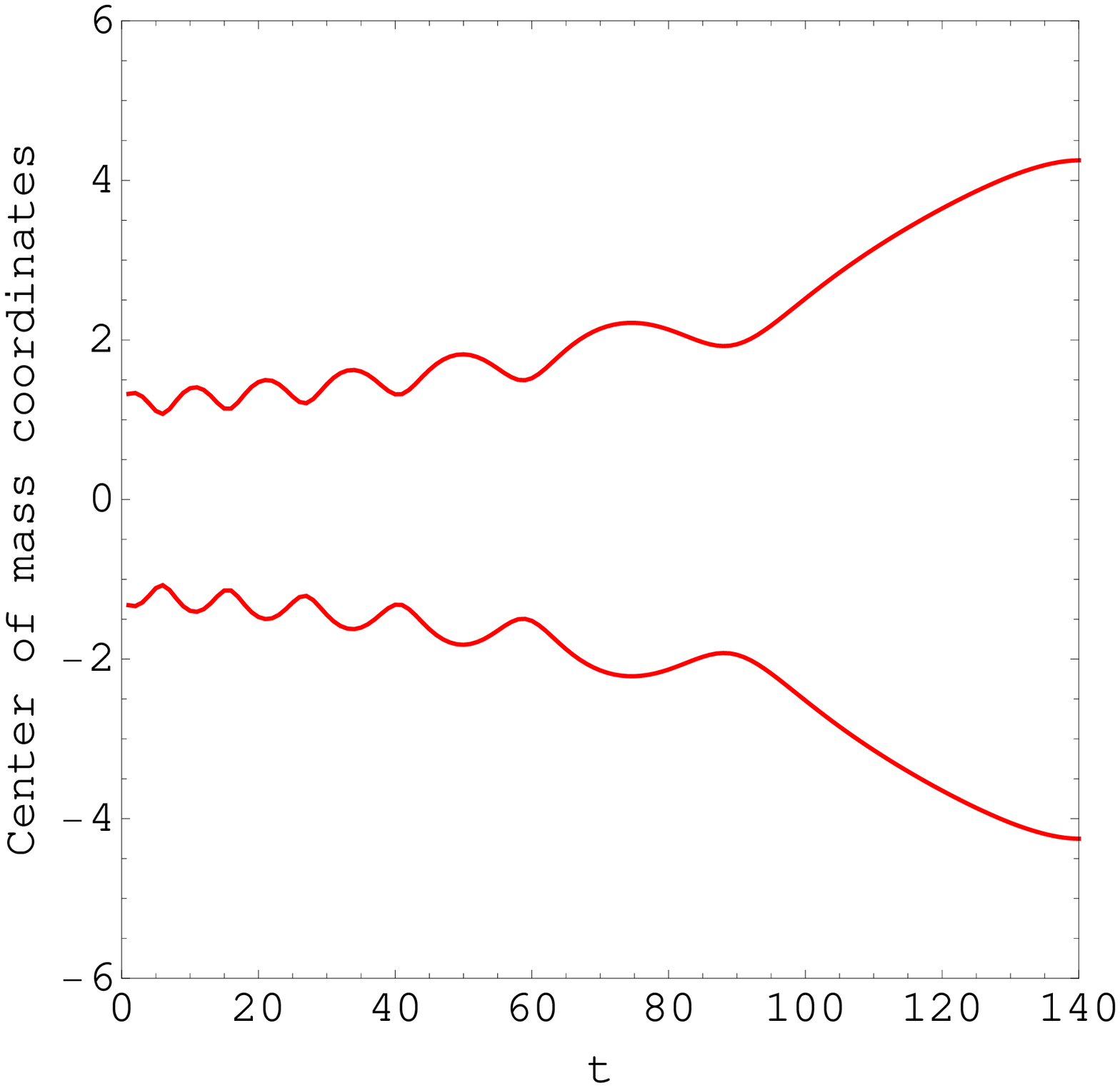}}
\caption{Left panel: The waveforms for a strongly bound ($g=20$) and
weakly bound ($g=12$) two-soliton molecule, corresponding to the
fixed point of Eq. (\ref{att2}) with $N=2$, $q=-1$,
$\sigma=\sqrt{\pi}$. Right panel: Dynamics of center-of-mass
positions of solitons in the two-soliton molecule. When the strength
of attractive dipolar interactions is slowly decreased with time
$g(t)=g_0 (1-t/t_{end})$ for $g_0=12$, $t_{end}=200$ the distance
between solitons increases in oscillatory manner and the molecule
eventually disintegrates at $t \simeq 90$, corresponding to $g
\simeq 7$. The oscillations help to detect the time instance, when
the disintegration of the molecule occurs. The vibrations of the
molecule is induced by slight chirping of the initial waveform
$\psi(x,0) \, e^{-i b x^2}$, with $ b=0.05$. } \label{fig2}
\end{figure}

Substitution of Eq. (\ref{ansatz2}) into Eq. (\ref{lagden}) and
integration over space variable gives the effective Lagrangian

\begin{equation}
\frac{L}{N}=\frac{3}{2}a^2b_t-\frac{1}{2}\xi_t^2+\phi_t+
\frac{3}{4a^2}+3a^2b^2 - \frac{3\, q N}{8\sqrt{2\pi}a} -\frac{g
N}{16\sqrt{2\pi}} F_m(a,\sigma).
\end{equation}
The equation for the width can be obtained from the Euler-Lagrange
equations
\begin{equation}\label{att2}
a_{tt}=\frac{1}{a^3}-\frac{q N}{4\sqrt{2\pi}a^2}+\frac{g
N}{24\sqrt{2\pi}}\frac{\partial F_m(a,\sigma)}{\partial a},
\end{equation}
where
\begin{eqnarray}
F_m(a,\sigma)&=&\exp[(2\sigma a)^{-2}] \left([1+3(\sigma
a)^2+3(\sigma a)^4]
K_1[(2\sigma a)^{-2}] \right. \nonumber \\
& & \left. -[1+5(\sigma a)^2+7(\sigma a)^4]K_0[(2\sigma
a)^{-2}]\right)/(\sigma a)^7, \label{Fm}
\end{eqnarray}
$K_0(x), K_1(x)$ - are the modified Bessel functions of the second
kind. The fixed point of Eq. (\ref{att2}) gives the stationary shape
of the two-soliton molecule. The waveforms of a two-soliton molecule
according to these equations are presented in Fig. \ref{fig2} for
two different strengths of dipolar interactions. The center-of-mass
positions of the solitons are located at $x_0 = \pm \,
2a_0/\sqrt{\pi}$, where $a_0$ is the fixed point of Eq.
(\ref{att2}).

The properties of two-soliton molecules in the weakly bound regime
can be probed by slowly reducing the strength of the dipolar atomic
interactions. In the experiments tuning the dipolar interactions can
be achieved by fast rotation of the orientation of the dipoles in a
magnetic field \cite{giovanazzi2002,tang2018}. When the competing
long-range dipolar and short-range contact interactions nearly
balance each-other, the bound state of two solitons becomes unstable
against small perturbations. A slight imbalance may lead to
disintegration of the molecule, as shown on the right panel of Fig.
\ref{fig2}. In numerical simulations we induce mild vibrations of
the molecule by chirping the initial waveform. The oscillatory
component in the dynamics helps to detect precisely the time
instance when the dissociation of the molecule occurs. Just before
the disintegration the soliton molecule has a very small binding
energy. Hence, without the oscillatory component in the dynamics, it
would be difficult to detect the time instance, when the molecule
dissociates into freely moving solitons.

\section{Potential of interaction between solitons}

\begin{figure}[htb]
\centerline{
\includegraphics[width=6cm,height=6cm,clip]{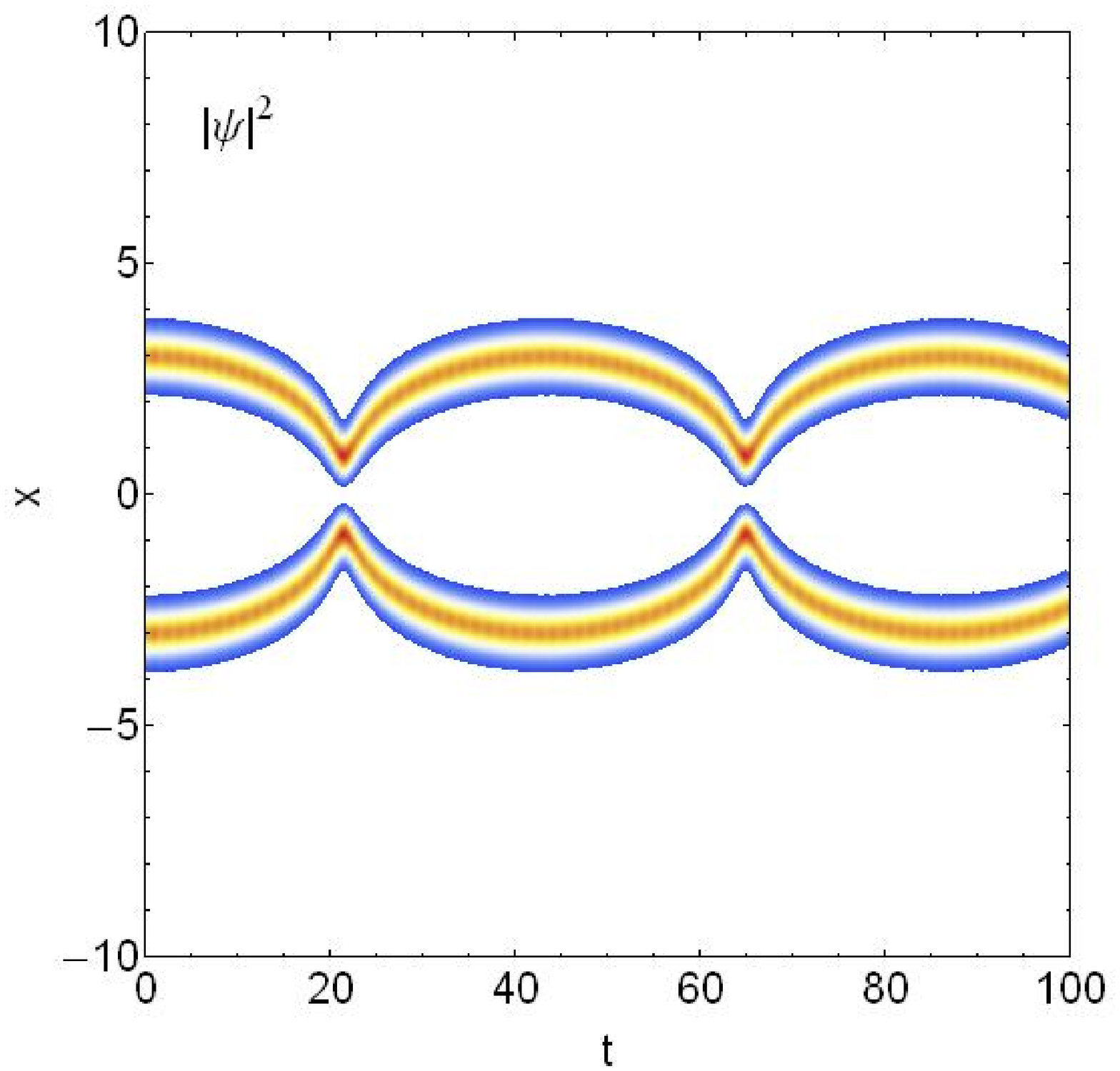} \quad\\
\includegraphics[width=6cm,height=6cm,clip]{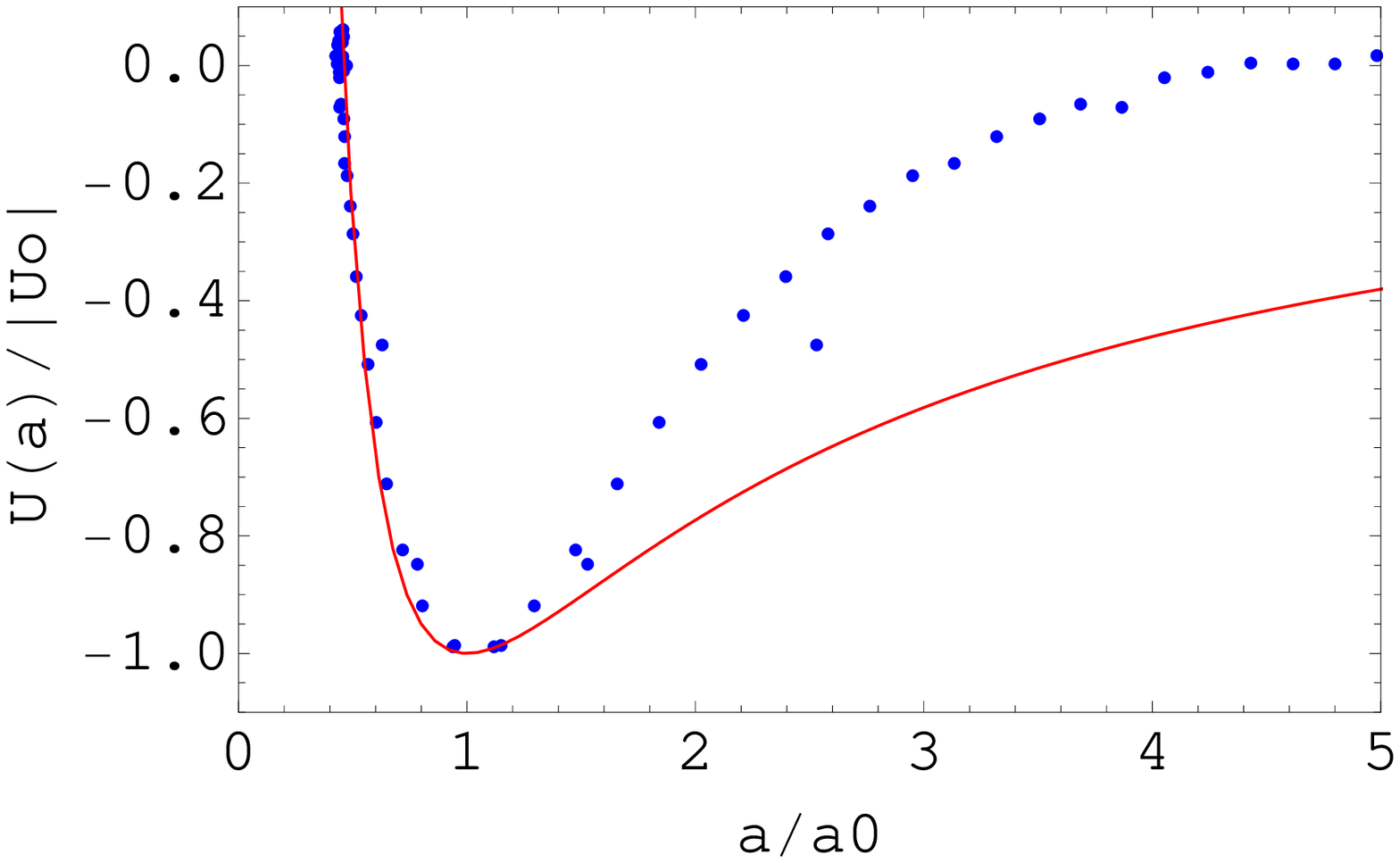}}
\caption{Left panel: If two solitons are initially placed at a
distance larger than the equilibrium separation ($x_0=\pm
\,2a_0\sqrt{\pi}$) they attract each - other due to the long-range
dipolar interactions. When the inter-soliton distance becomes small,
repulsion due to anti-phase configuration becomes prevailing and
solitons start to move in opposite direction. This process gives
rise to oscillations of the two-soliton molecule. Right panel: The
potential of interaction between two solitons according to VA (red
solid line) and numerical solution of the nonlocal GPE (\ref{gpe})
without external potential (blue symbols). } \label{fig3}
\end{figure}

From numerical simulations of the GPE (\ref{gpe}) it was noticed,
that the potential of interaction between solitons at large
separations ($a~\gg~a_0$) notably deviate from the prediction of
variational approach (see the right panel of Fig. \ref{fig3}).
Obviously, this is due to the fact that the ansatz (\ref{ansatz2})
does not well represent two separate solitons when the distance
between them is large.

The potential of interaction between two solitons, shown on the
right panel of  Fig. \ref{fig3} by blue symbols, has been
constructed as follows: Two sufficiently far separated solitons (but
still ``feeling" each - other) are prepared in anti-phase
configuration as initial condition in Eq. (\ref{gpe}). Due to the
long - range dipolar attractive interaction they start to move
toward each - other. When the inter-soliton distance becomes small,
they repel due to the anti-phase configuration. The maximum and
minimum of the center of mass positions correspond to turning points
$x_{p_{1,2}}$ in the potential well. Each time when the soliton
passes its equilibrium position ($x=\pm \, 2a_0/\sqrt{\pi}$), the
potential energy fully transforms into kinetic energy. By measuring
the maximal velocity $v_{m}$ one can define the potential energy at
the turning points $U(x_{p_{1,2}}) = v_{m}^2/2$, where the mass of
the soliton is set to one. Repeating this procedure with different
initial separations, one can construct the shape of the potential.

\begin{figure}[htb]
\centerline{
\includegraphics[width=8cm,height=6cm,clip]{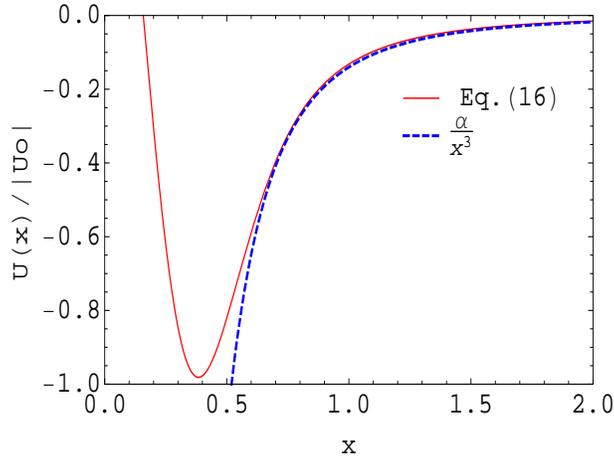}}
\caption{The potential Eq.(\ref{pot2}) originating from the presence
of one soliton. The other soliton moves under the influence of this
potential. At long distances the potential decays as $\sim 1/x^3$,
which is a characteristic of dipolar interactions. The fitting
parameter is equal to $\alpha = - 0.14$.} \label{fig4}
\end{figure}

An alternative approach to evaluation of the interaction potential
between two solitons consists in analyzing the equation for the
center-of-mass $x_c$ of one soliton moving in the potential created
by the other soliton \cite{dai2016}. Suppose we have ``left" and
``right" solitons positioned at $x_c$ and $-x_c$ respectively. In
the case of attractive interactions the left soliton moves towards
the right soliton due to its modification of the ``refractive index"
of the medium
\begin{eqnarray}
\frac{d\,^2 x_c}{d t^2} &=& -\frac{d}{dx}\left(\Delta n_R \right)\mid_{x=-x_c}, \\
\Delta n_R &=& q|\psi(x,t)|^2 + g\int
\limits_{-\infty}^{\infty}R(|x-x'|)|\psi(x',t)|^2 dx'\nonumber
\end{eqnarray}
Using the response function (\ref{kernel}) and ansatz
(\ref{ansatz1}) we obtain the equation of motion for the
center-of-mass coordinate of the left soliton
\begin{equation}
\frac{d^{\,2} x_c}{d t^2} = -\frac{4qN}{a^3 \sqrt{\pi}}\, x_c \,
e^{-4x_c^2/a^2} - \frac{3 g N \sigma^2}{a\sqrt{\pi}} \int
\limits_{-\infty}^{\infty} \frac{t \, e^{-(t-2x_c)^2/a^2}}{(\sigma^2
t^2+1)^{5/2}}dt.
\end{equation}
On the right hand side of this equation we have the force, acting on
the left soliton. The potential is the integral of the force with
respect to $x_c$ (below we write it by dropping the subscript)
\begin{equation}\label{pot2}
U(x)=\frac{q N}{2\sqrt{\pi} a} e^{-4x^2/a^2} + \frac{3 g N
\sigma^2}{4} \int \limits_{-\infty}^{\infty} \frac{\tau \,{\rm
erfc}[(\tau-2x)/a]}{(\sigma^2 \tau^2 + 1)^{5/2}} d \tau,
\end{equation}
where ${\rm erfc}(z)$ is the complimentary error
function\cite{abramowitz}. The first term with $q<0$ is responsible
for the short-range repulsive interaction, while the second term
with $g>0$ corresponds to the long-range attractive interaction. The
potential Eq. (\ref{pot2}) is depicted in Fig. \ref{fig4}.
By comparing this figure with the right panel of Fig. \ref{fig3} one
can observe the qualitative agreement between the numerically
constructed potential and the prediction of Eq. (\ref{pot2}).

\section{The binding energy of a two-soliton molecule}

The total energy of a single soliton or two-soliton molecule can be
evaluated from the GPE energy functional
\begin{equation}\label{energy}
E=\int \limits_{-\infty}^{\infty}\left[\frac{1}{2}|\psi_x|^2 +
\frac{q}{2}|\psi|^4 + \frac{g}{2}|\psi|^2 \int
\limits_{-\infty}^{\infty} R(|x-x'|) |\psi(x')|^2 dx' \right]dx,
\end{equation}
where $\psi(x)$ is the waveform of the localized state, while $R(x)$
is the response function.

By substituting in Eq. (\ref{energy}) the waveforms, corresponding
to the single soliton Eq. (\ref{ansatz1}) and two-soliton molecule
Eq. (\ref{ansatz2}) we get the corresponding energies
\begin{equation}
E_s = \frac{N_s}{4a_s^2}-\frac{q N_s^2}{2\sqrt{2\pi}a_s}-\frac{g
N_s^2}{2\sqrt{2}\sigma} F_s(a_s,\sigma),
\end{equation}
where $N_s$, $a_s$ - are the norm and width of a single soliton,
respectively, while the function $F_s(a_s,\sigma)$ is given by Eq.
(\ref{Fs}).

Similar expression for the total energy of the molecule has the form
\begin{equation}
E_m = \frac{3N_m}{4a_m^2}-\frac{3\, q
N_m^2}{8\sqrt{2\pi}a_m}-\frac{g N_m^2}{16\sqrt{2\pi}}F(a_m,\sigma),
\end{equation}
where $N_m$, $a_m$ - are the norm and width of a two-soliton
molecule, respectively, while the function $F_m(a_m,\sigma)$ is
given by Eq. (\ref{Fm}).

The binding energy of a two-soliton molecule can be estimated as the
difference between the energy of the molecule, and the energies of
two solitons well separated from each other, i.e. non-interacting
solitons.
\begin{equation}
\Delta E = E_m - 2 E_s.
\end{equation}
To simplify the resulting expression, we use the approximate
relations between the parameters of the molecule and single solitons
which form this molecule being placed at the equilibrium distance.

\begin{figure}[htb]
\centerline{\includegraphics[width=6cm,height=6cm,clip]{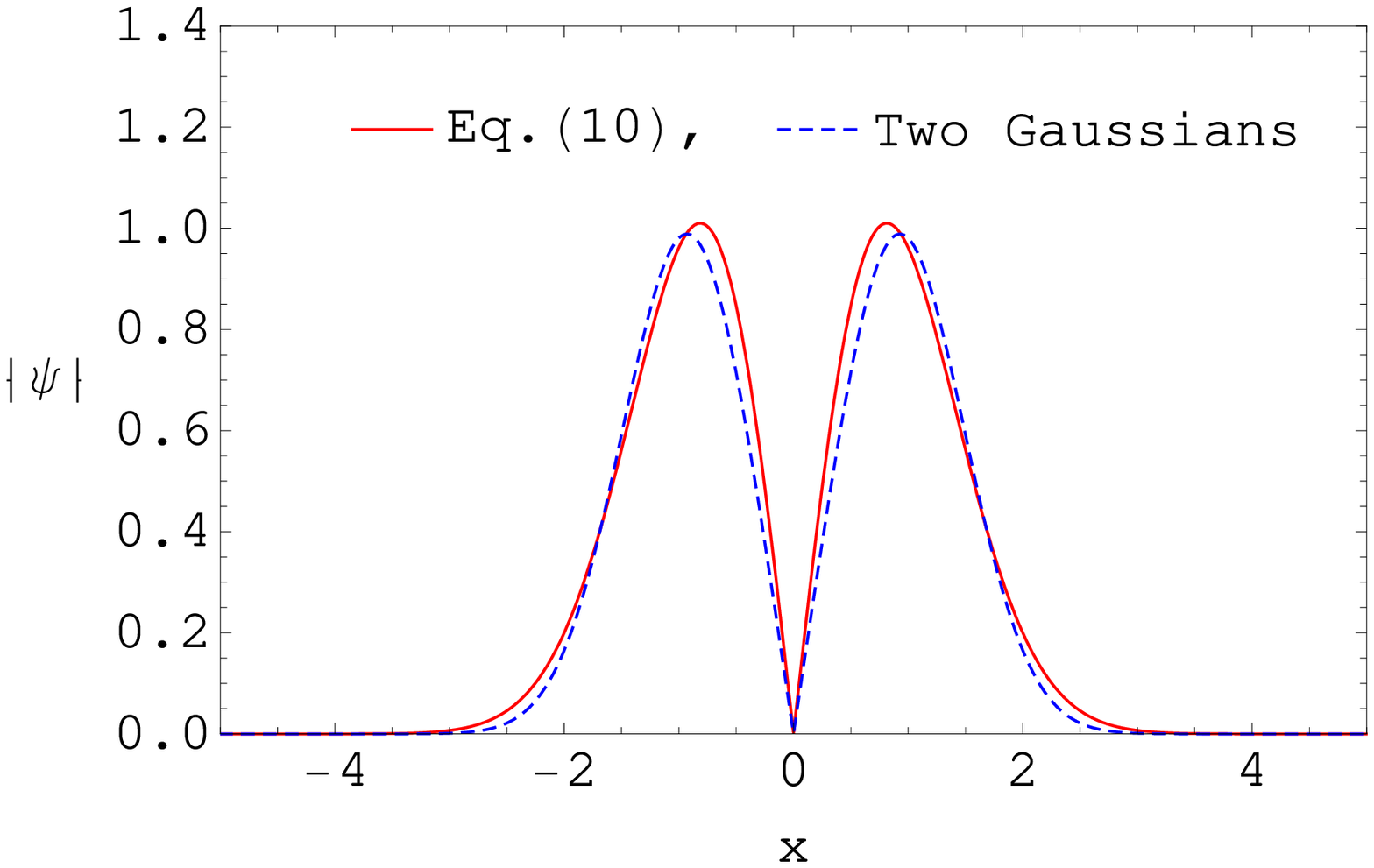}\quad
            \includegraphics[width=6cm,height=6cm,clip]{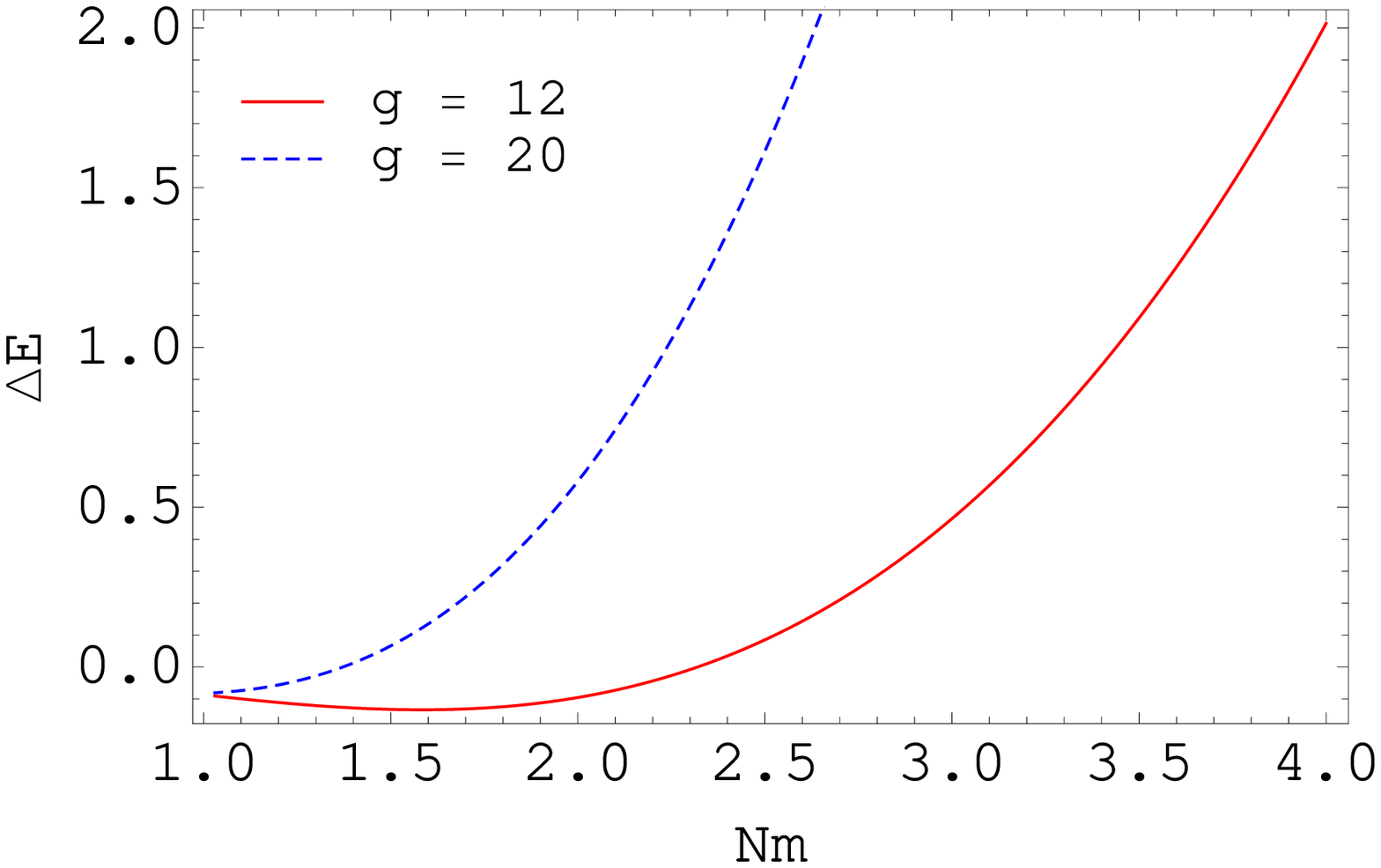}}
\caption{Left panel: The waveform Eq.(\ref{ansatz2}) as
approximation by two Gaussian functions Eq. (\ref{Aa}) separated by
a distance $x_0=1.82$,. Parameter values: $A_m=2.04$, $a_m=0.81$,
$N=2$. Right panel: The binding energy of the two-soliton molecule
as a function of its norm for different strengths of the dipolar
interactions. Weakly bound soliton molecule with $g=12$
disintegrates ($\Delta E \rightarrow 0$) even though its norm is
sufficiently large $N_m \simeq 2.5$.} \label{fig5}
\end{figure}

The waveform of a two-soliton molecule (\ref{ansatz2}) can be well
approximated by two anti-phase solitons, as illustrated in
Fig.~\ref{fig5}
\begin{equation}\label{init}
\psi(x) = A_s \left[e^{-(x-x_0)^2/2a_s^2}-e^{-(x+x_0)^2/2a_s^2}
\right],
\end{equation}
with the following amplitude, width and half separation
\cite{baizakov2015}
\begin{equation}\label{Aa}
A_s = \frac{2 A_m a_m}{\sqrt{\pi}} e^{-2/\pi}, \quad a_s = \frac{\pi
a_m}{16} e^{4/\pi}, \quad x_0 = \frac{2 a_m}{\sqrt{\pi}}.
\end{equation}

Due to the destructive interference the norm of the molecule will be
slightly less than twice of the soliton's norm. We shall neglect
this effect and set $N_s = N_m/2$. These relations lead to the
expression for the binding energy of the molecule
\begin{eqnarray}
\Delta E &=& \frac{N_m}{4 \pi^2 a_m^2}\left[3\pi^2-512\, e^{-8/\pi}
\right]-\frac{qN_m^2}{8\sqrt{2}\pi^{3/2} a_m} \left[3\pi-32\, e^{-4/\pi} \right] - \nonumber \\
& & \frac{g N_m^2}{24\sqrt{\pi}\sigma} \left[4\sigma F_m -
3\sqrt{2\pi} F_s \right].
\end{eqnarray}
where $N_m$, $a_m$ are the norm and equilibrium width of the
molecule, given by the fixed point of Eq. (\ref{att2}).

\section{Conclusions}

The analytic expressions for the potential of interaction between
bright solitons and the binding energy of a two soliton molecule in
quasi-1D dipolar BEC have been derived. The mathematical model,
based on the variational approach, has been corroborated with the
results of numerical simulations of the nonlocal Gross-Pitaevskii
equation. The developed model allows to explore the properties of
solitons and two-soliton molecules in the weakly bound regime, when
the short-range contact atomic interactions and long-range
dipole-dipole interactions nearly balance each-other. The obtained
results can be of interest in future experiments with dipolar BEC,
including the conditions close to the formation of quantum droplets.

\section*{Acknowledgments}
This work has been supported by the KFUPM research group projects
RG1503-1 and RG1503-2. B.B.B and U.A.K thank the Physics Department
at KFUPM and Saudi Center for Theoretical Physics for the
hospitality provided during their visits. \\


\begin{thebibliography}{99}

\bibitem{griesmaier2005}
A. Griesmaier, J. Werner, S. Hensler, J. Stuhler, T. Pfau,
Bose-Einstein condensation of chromium, Phys. Rev. Lett.  {\bf 94}
160401 (2005).

\bibitem{beaufils2008}
Q. Beaufils, R. Chicireanu, T. Zanon, B. Laburthe-Tolra, E.
Mar\'echal, L. Vernac, J. C. Keller, and O. Gorceix, All-optical
production of chromium Bose-Einstein condensates, Phys. Rev. A {\bf
77}, 061601(R) (2008).

\bibitem{lu2011}
M. Lu, N. Q. Burdick, S. H. Youn, B. L. Lev, Strongly dipolar
Bose-Einstein condensate of dysprosium, Phys. Rev. Lett. {\bf 107},
190401 (2011).

\bibitem{tang2015}
Y. Tang, N. Q. Burdick, K. Baumann, B. L. Lev, Bose-Einstein
condensation of $^{162}$Dy and $^{160}$Dy, New J. Phys. {\bf 17},
045006 (2015 ).

\bibitem{aikawa2012}
K. Aikawa, A. Frisch, M. Mark, S. Baier, A. Rietzler, R. Grimm, F.
Ferlaino, Bose-Einstein condensation of erbium, Phys. Rev. Lett.
{\bf 108}, 210401 (2012).

\bibitem{trautmann2018}
A. Trautmann, P. Ilzh\"ofer, G. Durastante, C. Politi, M. Sohmen, M.
J. Mark, and F. Ferlaino, Dipolar quantum mixtures of erbium and
dysprosium atoms, Phys. Rev. Lett. {\bf 121}, 213601 (2018).

\bibitem{lahaye2009}
T. Lahaye, C. Menotti, L. Santos, M. Lewenstein, T. Pfau, The
physics of dipolar bosonic quantum gases, Rep. Prog. Phys. {\bf 72},
 126401 (2009).

\bibitem{pethick-book}
C. J. Pethick, H. Smith, Bose-Einstein condensation in dilute gases,
Cambridge University Press, Cambridge, 2002.

\bibitem{pitaevskii-book}
L. Pitaevskii, S. Stringari, Bose-Einstein condensation, Oxford
Uniersity Press, Oxford, 2003.

\bibitem{lahaye2008}
T. Lahaye, J. Metz, B. Fr\"ohlich, T. Koch, M. Meister, A.
Griesmaier, T. Pfau, H. Saito, Y. Kawaguchi, M. Ueda, d-Wave
collapse and explosion of a dipolar Bose-Einstein condensate, Phys.
Rev. Lett. {\bf 101}, 080401 (2008).

\bibitem{santos2003}
L. Santos, G. V. Shlyapnikov, M. Lewenstein, Roton-Maxon spectrum
and stability of trapped dipolar Bose-Einstein condensates, Phys.
Rev. Lett. {\bf 90}, 250403 (2003).

\bibitem{petter2018}
D. Petter, G. Natale, R. M. W. van Bijnen, A. Patscheider, M. J.
Mark, L. Chomaz, and F. Ferlaino, Probing the roton excitation
spectrum of a stable dipolar Bose gas, arXiv preprint
arXiv:1811.12115 (2018).

\bibitem{wilson2010}
R. M. Wilson, S. Ronen, J. L. Bohn, Dipolar Bose-Einstein
condensates as discrete superfluids, Phys. Rev. Lett. {\bf 104},
094501 (2010).

\bibitem{ticknor2011}
C. Ticknor, R. M. Wilson, J. L. Bohn, Anisotropic superfluidity in a
dipolar Bose gas, Phys. Rev. Lett.  {\bf 106}, 065301 (2011).

\bibitem{zaman2010}
M. Asad-uz Zaman, D. Blume, Aligned dipolar Bose-Einstein condensate
in a double-well potential: From cigar-shaped to pancake-shaped, New
J. Phys. {\bf 12}, 065022 (2010).

\bibitem{nath2010}
R. Nath, L. Santos, Faraday patterns in dipolar Bose-Einstein
condensates, Phys. Rev. A {\bf 81}, 033626 (2010).

\bibitem{baranov2008}
M. A. Baranov, Theoretical progress in many-body physics with
ultracold dipolar gases, Phys. Rep.  {\bf 464}, 71 (2008).

\bibitem{yukalov2016}
V. I. Yukalov, E. P. Yukalova, Bose-condensed atomic systems with
nonlocal interaction potentials, Laser Phys. {\bf 26}, 045501
(2016).

\bibitem{baillie2015}
D. Baillie, P. B. Blakie, A general theory of flattened dipolar
condensates, New J. Phys.  {\bf 17}, 033028 (2015).

\bibitem{pedri2005}
P. Pedri, L. Santos, Two-Dimensional bright solitons in dipolar
Bose-Einstein condensates, Phys. Rev. Lett.  {\bf 95}, 200404
(2005).

\bibitem{tikhonenkov2008}
I. Tikhonenkov, B. A. Malomed, A. Vardi, Anisotropic solitons in
dipolar Bose-Einstein condensates, Phys. Rev. Lett.  {\bf 100},
090406 (2008).

\bibitem{tikhonenkov2008a}
I. Tikhonenkov, B. A. Malomed, A. Vardi, Vortex solitons in dipolar
Bose-Einstein condensates, Phys. Rev.  A {\bf 78}, 043614 (2008).

\bibitem{cuevas2009}
J. Cuevas, B. A. Malomed, P. G. Kevrekidis, D. J. Frantzeskakis,
Solitons in quasi-one-dimensional Bose-Einstein condensates with
competing dipolar and local interactions, Phys. Rev. A {\bf 79},
 053608 (2009).

\bibitem{young2011}
L. E. Young-S, P. Muruganandam, S. K. Adhikari, Dynamics of
quasi-one-dimensional bright and vortex solitons of a dipolar
Bose-Einstein condensate with repulsive atomic interaction, J. Phys.
B: At. Mol. Opt. Phys. {\bf 44}, 101001 (2011).

\bibitem{eichler2012}
R. Eichler, D. Zajec, P. K\"oberle, J. Main, G. Wunner, Collisions
of anisotropic two-dimensional bright solitons in dipolar
Bose-Einstein condensates, Phys. Rev.  A {\bf 86}, 053611 (2012).

\bibitem{umarov2016}
B. A. Umarov, N. A. B. Aklan, B. B. Baizakov and F. Kh. Abdullaev,
Scattering of a two-soliton molecule by Gaussian potentials in
dipolar Bose-Einstein condensates, J. Phys. B: At. Mol. Opt. Phys.
{\bf 49}, 125307 (2016).

\bibitem{adhikari2014}
S. K. Adhikari, Bright dipolar Bose-Einstein condensate soliton
mobile in a direction perpendicular to polarization, Phys. Rev. A
{\bf 89}, 043615 (2014).

\bibitem{abdullaev2012}
F. Kh. Abdullaev, V. A. Brazhnyi, Solitons in dipolar Bose-Einstein
condensates with trap and barrier potential, J. Phys. B: At. Mol.
Opt. Phys. {\bf 45}, 085301 (2012).

\bibitem{edmonds2017}
M. J. Edmonds, T. Bland, R. Doran and N. G. Parker, Engineering
bright matter-wave solitons of dipolar condensates, New J. Phys.
{\bf 19}, 023019 (2017).

\bibitem{bland2017}
T. Bland, K. Pawlowski, M. J. Edmonds, K. Rzazewski, and N. G.
Parker, Interaction-sensitive oscillations of dark solitons in
trapped dipolar condensates, Phys. Rev. A {\bf 95}, 063622 (2017).

\bibitem{nath2007}
R. Nath, P. Pedri, L. Santos, Soliton-soliton scattering in dipolar
Bose-Einstein condensates, Phys. Rev. A {\bf 76}, 013606 (2007).

\bibitem{lakomy2012}
K. Lakomy, R. Nath, L. Santos, Soliton molecules in dipolar
Bose-Einstein condensates, Phys. Rev. A {\bf 86}, 013610 (2012).

\bibitem{lashkin2007}
V. M. Lashkin, Two-dimensional nonlocal vortices, multipole
solitons, and rotating multi solitons in dipolar Bose-Einstein
condensates, Phys. Rev. A {\bf 75}, 043607 (2007).

\bibitem{turmanov2015}
B. Kh. Turmanov, B. B. Baizakov, B. A. Umarov, F. Kh. Abdullaev,
Vibration spectrum of a two-soliton molecule in dipolar
Bose-Einstein condensates, Phys. Lett. A {\bf 379}, 1828 (2015).

\bibitem{baizakov2015}
B. B. Baizakov, S. M. Al-Marzoug, H. Bahlouli, Interaction of
solitons in one-dimensional dipolar Bose-Einstein condensates and
formation of soliton molecules, Phys. Rev. A {\bf 92}, 033605
(2015).

\bibitem{pawlowski2015}
K. Pawlowski, K. Rzazewski, Dipolar dark solitons, New J. Phys. {\bf
17}, 105006 (2015).

\bibitem{bland2015}
T. Bland, M. J. Edmonds, N. P. Proukakis, A. M. Martin, D. H. J.
O'Dell, N. G. Parker, Controllable non-local interactions between
dark solitons in dipolar condensates, Phys. Rev. A {\bf 92}, 063601
(2015).

\bibitem{ferrier-barbut2016}
I. Ferrier-Barbut, H. Kadau, M. Schmitt, M. Wenzel, and T. Pfau,
Observation of quantum droplets in a strongly dipolar Bose gas,
Phys. Rev. Lett. {\bf 116}, 215301 (2016).

\bibitem{kadau2016}
H. Kadau, M. Schmitt, M. Wenzel, C. Wink, T. Maier, I. Ferrier-
Barbut, and T. Pfau, Observing the Rosensweig instability of a
quantum ferrofluid, Nature {\bf 530}, 194 (2016).

\bibitem{schmitt2016}
M. Schmitt, M. Wenzel, F. B\"ottcher, I. Ferrier-Barbut and T. Pfau,
Self-bound droplets of a dilute magnetic quantum liquid, Nature {\bf
539}, 259 (2016).

\bibitem{carr2009}
L. D. Carr, D. DeMille, R. V. Krems, J. Ye, Cold and ultracold
molecules: Science, technology, and applications, New J. Phys. {\bf
11}, 055049 (2009).

\bibitem{helm2015}
J. L. Helm, S. L. Cornish, S. A. Gardiner, Sagnac interferometry
using bright matter-wave solitons, Phys. Rev. Lett. {\bf 114},
 134101 (2015).

\bibitem{sapina2013}
I. Sapina, T. Dahm, Interaction of a Bose-Einstein condensate and a
superconductor via eddy currents, New J. Phys. {\bf 15}, 073035
(2013).

\bibitem{sapina2014}
I. Sapina, T. Dahm, Dynamics of a dipolar Bose-Einstein condensate
in the vicinity of a superconductor, Phys. Rev. A {\bf 90}, 052709
(2014).

\bibitem{lima2011}
A. R. P. Lima and A. Pelster, Quantum fluctuations in dipolar Bose
gases, Phys. Rev. A {\bf 84}, 041604(R) (2011).

\bibitem{edler2017}
D. Edler, C. Mishra, F. W\"achtler, R. Nath, S. Sinha, and L.
Santos, Quantum fluctuations in quasi-one-dimensional dipolar
Bose-Einstein condensates, Phys. Rev. Lett. {\bf 119}, 050403
(2017).

\bibitem{sinha2007}
S. Sinha, L. Santos, Cold dipolar gases in quasi-one-dimensional
geometries, Phys. Rev. Lett. {\bf 99},  140406 (2007).

\bibitem{anderson1983}
D. Anderson, Variational approach to nonlinear pulse propagation in
optical fibers, Phys. Rev. A {\bf 27}, 1393 (1983).

\bibitem{malomed2002}
B. A. Malomed, Variational methods in nonlinear fiber optics and
related fields, in: E. Wolf (Ed.), Progr. Opt., vol. {\bf 43},
North-Holland, Amsterdam, 2002, pp.69 - 191.

\bibitem{abramowitz}
M. Abramowitz, I. A. Stegun, Handbook of mathematical functions,
National Bureau of Standards, Washington, 1964.

\bibitem{pare1999} C. Pare and P. A. Belanger, Antisymmetric soliton in a
dispersion-managed system, Optics Commun. {\bf 168}, 103-109 (1999).

\bibitem{feng2004}
B. F. Feng and B. A. Malomed, Antisymmetric solitons and their
interactions in strongly dispersion-managed fiber-optic systems,
Optics. Commun. {\bf 229}, 173-185 (2004).

\bibitem{giovanazzi2002}
S. Giovanazzi, A. G\"orlitz, and T. Pfau, Tuning the dipolar
interaction in quantum gases, Phys. Rev. Lett. {\bf 89}, 130401
(2002).

\bibitem{tang2018}
Y. Tang, W. Kao, K.-Y. Li, and B. L. Lev, Tuning the dipole- dipole
interaction in a quantum gas with a rotating magnetic field, Phys.
Rev. Lett. {\bf 120}, 230401 (2018).

\bibitem{dai2016}
Zhiping Dai, Zhenjun Yang, Xiaohui Ling, Shumin Zhang, Zhaoguang
Pang, Interaction trajectory of solitons in nonlinear media with an
arbitrary degree of nonlocality, Ann. Phys. {\bf 366}, 13 (2016)

\end{thebibliography}
\end{document}